\documentclass[prc,aps,twocolumn,floats,floatfix,showpacs,superscriptaddress,nofootinbib]{revtex4}
\usepackage{amsmath}
\usepackage{amsfonts}
\usepackage{graphicx}

\usepackage{color}

\newcommand{\bo}{\boldsymbol}

\begin{document}

\title{Sensitivity of $\beta$-decay rates to the radial dependence of the nucleon effective mass}

\author{A. P. Severyukhin}
\affiliation{Bogoliubov Laboratory of Theoretical Physics,
             Joint Institute for Nuclear Research,
             141980 Dubna, Moscow region, Russia}
\author{J. Margueron}
\affiliation{Institut de Physique Nucl\'eaire de Lyon, Universit\'e Claude Bernard Lyon 1, \\IN2P3-CNRS, F-69622 Villeurbanne}
\author{I. N. Borzov}
\affiliation{Bogoliubov Laboratory of Theoretical Physics,
             Joint Institute for Nuclear Research,
             141980 Dubna, Moscow region, Russia}
\author{N. Van Giai}
\affiliation{Institut de Physique Nucl\'eaire, CNRS-IN2P3 and
             Univ. Paris-Sud, 91405 Orsay, France}

\begin{abstract}
We analyze the sensitivity of $\beta$-decay rates in $^{78}$Ni and
$^{100,132}$Sn to a correction term in Skyrme energy-density
functionals (EDF) which modifies the radial shape of the nucleon
effective mass. This correction is added on top of several Skyrme
parametrizations which are selected from their effective mass
properties and predictions about the stability properties of
$^{132}$Sn. The impact of the correction on high-energy collective
modes is shown to be moderate. From the comparison of the effects
induced by the surface-peaked effective mass in the three doubly
magic nuclei, it is found that $^{132}$Sn is largely impacted by
the correction, while $^{78}$Ni and $^{100}$Sn are only moderately
affected. We conclude that $\beta$-decay rates in these nuclei can
be used as a test of different parts of the nuclear EDF: $^{78}$Ni
and $^{100}$Sn are mostly sensitive to the particle-hole
interaction through the B(GT) values, while $^{132}$Sn is
sensitive to the radial shape of the effective mass. Possible
improvements of these different parts could therefore be better
constrained in the future.
\end{abstract}


%\date{}
\date{\today}

\maketitle
%
%=======================================================
%
\section{Introduction}

Weak processes such as $\beta$-decay rates, electron capture,
neutrino scattering and absorption play an important role during
the late evolution of massive stars~\cite{Bethe:1990}. They are
largely responsible for the electron fraction in the core during
the core-collapse phase and they continue to play a determinant
role in the nuclear synthesis $r$-process~\cite{Langanke:2003}.
Because of their great importance in astrophysical applications,
weak processes were extensively investigated within various
approaches. The large-scale shell model Monte Carlo (SMMC) method
was, for instance, applied to compute $\beta^\pm$ decay rates for
stellar conditions for more than 100 nuclei in the mass range A =
45-65~\cite{Langanke:2000,Pinedo:2000}. Recently, mean-field based
models have been used for the prediction of electron-capture cross
sections and rates. Finite-temperature charge-exchange RPA (CERPA)
models based on Skyrme or relativistic functionals have been
applied to predict electron-capture cross sections using different
interactions~\cite{Paar:2009,Niu:2011,Fantina:2012}. Mean-field
predictions around the Fermi energy are, however, known to suffer
from their underestimation of the density of states. In this
work, we explore a small correction to the mean field models which
increases the density of states around the Fermi
energy~\cite{Fantina:2011}. Here, we compare the predictions of
this model to known experimental values such as $\beta$ half-lives
or collective modes, as a first step before using it for
astrophysical applications.

Since the pioneering work of Brown et al.~\cite{Brown:1964} it is
known that the level density around the Fermi energy in stable
nuclei indicates that the in-medium nucleon effective mass is
close to the bare mass. The description of giant resonances such
as the giant dipole resonance requires, on the other hand,
that the nucleon effective mass in the nuclear medium should be
reduced as compared to its value in
vacuum~\cite{Book:Harakeh:2001}. Analysis of the momentum
dependence of the nuclear optical potential also favors an
in-medium effective mass lower than in vacuum~\cite{JLM:1976,
MS:1991}.

These apparently diverging properties of the in-medium effective mass $m^{*}$ can be
reconciled by considering the two contributions to $m^{*}$:
the $k$-mass which is also called the non-locality mass, and the
$\omega$-mass which is induced by dynamical correlations such as
particle-phonon coupling~\cite{VMB:1976, BG:1980, CCSB:2014,Ma:1983,LA11}.
The coupling of the collective modes to the single-particle (s.p.) motion is,
however, difficult to perform in a self-consistent
approach. One of the main problems is coming from the
fragmentation of s.p. strength which increases exponentially at
each iteration of the self-consistent method.
{It has therefore been tried to include these correlation directly in the mean field,
either at the level of the interaction with density-dependent gradient terms~\cite{FPT97},
or, loosing the relation with an interaction, at the level of the nuclear energy density
functional (EDF) so as to produce a surface-peaked
effective mass (SPEM) which, at the same time, does not strongly
modify the mean-field~\cite{Fantina:2011}.
In this study, we will explore the second approach.}
%\blue{Alternative scheme to generate the SPEM was
%considered in Ref.~\cite{FPT97}.}

Predictions of $\beta$-decay rates throughout the nuclear chart
within a consistent microscopic nuclear model are difficult.
Tuning of models according to the system under study is usually
performed, and the description of $\beta$-decay rates through a
unique microscopic nuclear model does not exist. Since
$\beta$-decay rates are known to depend strongly on the fine
structure around the Fermi level, the difficulties to have a
general description could be related to the common issue with
mean-field models that the s.p. level density around the Fermi
level is too low. The increase of the level density, by using for
instance a model producing a SPEM could, in principle, lead to a
better description of $\beta$-decay rates throughout the nuclear
chart.

In microscopic approaches, calculations of nuclear $\beta$ -decay
rates are rather complex. Due to phase-space amplification
effects, the calculated $\beta$-decay rates are sensitive to both
nuclear binding energies and $\beta$-strength functions. In an
appropriate $\beta$-decay model, the correct amount of the
integral $\beta$-strength should be placed within the properly
calculated $Q_{\beta}$-window provided that the  spectral
distribution is also close to the "true" $\beta$-strength
function.  Furthermore, for consistency the model has to yield
correct positions and strengths of the Gamow-Teller (GT) and
first-forbidden resonances in the continuum~\cite{INB1}. Another
complication is related with the large-scale predictions of
nuclear $\beta$-decay rates. Such a program is a compromise
between accurate results and the necessity to cover extended
regions of the nuclear chart including deformed nuclei or even the
region with triple prolate-oblate-spherical shape coexistence
scenario. In this work, we shall consider only the case of
spherical nuclei. A plausible way to detect a change of the
$\beta$-decay strength function profile due to higher-order
corrections could be the analysis of the integral characteristics
of $\beta$-decay. The half-life is one of such characteristics,
being sensitive enough to the $\beta$-strength
distribution~\cite{INB1}. It is worth to analyze first the
doubly-magic $\beta^{\pm}$-unstable nuclides, such as
$^{100,132}$Sn, since one can use the simpler CERPA model. Also,
we compare to the most neutron-rich ($(N-Z)/A=0.28$) doubly-magic
nucleus $^{78}$Ni which is an important waiting point in the
r-process~\cite{H05}. The next step would then be to study the
delayed neutron and especially delayed multi-neutron
emission~\cite{Miernik}. This is a more difficult task since the
delayed neutron emission probability ($P_n$-value)~\cite{pn05}
puts an additional constraint on the $\beta$-strength distribution
in the near-threshold region.

This paper is organized as follows. In Sec.~II we briefly present
the modifications to the nuclear EDF which produce a SPEM, and we
describe the protocol used to adjust the strength of this
correction. In Sec.~III, we analyze the results of the
calculations of $\beta$-decay rates in $^{78}$Ni and
$^{100,132}$Sn and the properties of the giant quadrupole
resonance (GQR) and Gamow-Teller resonance (GTR) of $^{208}$Pb.
Conclusions are drawn in Sec.~IV.
%
%===============================================================
%
\section{The mean field models}
Skyrme-type EDF are known to give an accurate description of
masses and charge radii over the whole nuclear chart, from $Z=8$
up to heavy elements~\cite{Bender:2003}. As most of the mean-field
approaches, they however lead to a s.p. level density around the
Fermi surface which is lower than the experimental
one~\cite{Brown:1964}. Here, we introduce a correction term to the
Skyrme EDF which leads to a SPEM and increases the average s.p.
level density~\cite{Fantina:2011}. We hereafter present this
correction term and then briefly describe the calculations of
$\beta$-decay rates carried out consistently in the framework of
Hartree-Fock-CERPA approach.

\subsection{The standard Skyrme functional}
The standard Skyrme functional for the time-even energy density is
expressed as~\cite{Bender:2003}
\begin{eqnarray}
\mathcal{H}_{sky}(r) &=& \frac{\hbar^2}{2m}\tau_0+\sum_{t=0,1} C_t^\rho(\rho_0) \rho_t^2 + C_t^{\Delta \rho} \rho_t \Delta \rho_t
\nonumber \\
&&\hspace{0.5cm}+ C_t^\tau \rho_t\tau_t+\frac{1}{2}C_t^J J_t^2+C_t^{\nabla J} \rho_t\nabla\cdot J_t ,
\label{eq:functionalt}
\end{eqnarray}
where the indices $t=0,1$ stand for the isoscalar and isovector
part of the corresponding densities, respectively. For instance,
the nucleonic densities $\rho_0$ and $\rho_1$ are defined as,
\begin{eqnarray}
\rho_0(r) &=& \rho_n(r)+\rho_p(r),\nonumber \\
\rho_1(r) &=& \rho_n(r)-\rho_p(r),
\end{eqnarray}
where the densities $\rho_q$ ($q=n$, $p$) are expressed in terms
of the s.p. wave functions $\varphi_i^q$ as
\begin{equation}
\rho_q(r) = \sum_i \vert \varphi_i^q(r) \vert^2 .
\end{equation}
The kinetic energy and spin-current densities, $\tau_t$ and $J_t$,
are defined similarly~\cite{Bender:2003}. The coefficients $C_i^j$
in Eq.~(\ref{eq:functionalt}) are constants (see, e.g.,
Ref.~\cite{Bender:2003}) except for the coefficient $C_t^\rho$
which depends of the isoscalar density $\rho_0$ as:
\begin{equation}
C_t^\rho(\rho_0) = C_t^\rho(0) + (C_t^\rho(\rho_{0,sat})-C_t^\rho(0))\left(\frac{\rho_0}{\rho_{0,sat}}\right)^\alpha,
\end{equation}
where $\rho_{0,sat}$ is the saturation density in infinite nuclear matter.

The standard Skyrme functional can be separated into neutron and
proton channels, and neutron and proton effective masses
 are introduced:
\begin{equation}
\frac{m}{m_q^*} = 1 + \frac{2m}{\hbar^2}[(C_0^\tau+C_1^\tau)\rho_q+ (C_0^\tau-C_1^\tau)\rho_{\bar q} ] ,
\label{eq:effmassq}
\end{equation}
Then, neutron and proton mean fields can be obtained (see
appendix~\ref{app:npc}).

Among the large number of Skyrme parametrizations, we have selected
6 of them based on the following
requirements:
\begin{itemize}
\item First, the Skyrme EDF should
predict $^{132}$Sn as a $\beta$-unstable nucleus at the mean field
level. This is based on the common expectation that the Landau
parameter $G_0^\prime$ in the spin-isospin channel is repulsive
and will shift up the GT strength.
\item Second, we wish to explore different values of effective mass in the bulk, and different isospin splittings of the effective mass.
\end{itemize}

\begin{table*}[tb]
\caption{Bulk properties of the selected interactions.}
\label{tab:bulk}
\begin{ruledtabular}
\begin{tabular}{cccccccccc}
 Skyrme  & $\rho_{0,sat}$ & $E_0$ & $K_0$ & $J_{sym}$ & $L_{sym}$ & $m^*_s/m$ & $\Delta m^*/m$ & $G_0^\prime$ \\
   & (fm$^{-3}$) & (MeV) & (MeV) & (MeV) & (MeV) & (MeV) & & & \\
\noalign{\smallskip}\hline\noalign{\smallskip}
  SLyIII0.7~\cite{Washiyama:2012} &  0.153 & -16.33 & 361.4 & 31.98 & 30.78 & 0.7 &  0.18  & 0.30 \\
  SLyIII0.8~\cite{Washiyama:2012} &  0.153 & -16.32 & 368.8 & 31.69 & 28.24 & 0.8 &  0.29   & 0.33 \\
  SLyIII0.9~\cite{Washiyama:2012} &  0.153 & -16.31 & 374.5 & 31.44 & 24.75 & 0.9 &   0.38  & 0.34 \\
  $f_+$~\cite{Lesinski:2006}            &  0.162 & -16.04 & 230.0 & 32.00 & 41.53 & 0.7 &  0.17   & 0.08\\
  $f_0$~\cite{Lesinski:2006}            &  0.162 & -16.03 & 230.0 & 32.00 & 42.42 & 0.7 &  0        & -0.01\\
  $f_-$~\cite{Lesinski:2006}             &  0.162 & -16.02 & 230.0 & 32.00 & 43.79 & 0.7 & -0.28   & -0.13 \\
\end{tabular}
\end{ruledtabular}
\end{table*}

In $^{132}$Sn the first condition  can be related to the s.p.
energy difference between the $\pi 2d\frac 5 2$ and $\nu 2d\frac 3
2$ states (which contribute mostly to the GT transition towards
the $1^{+}$ state of $^{132}$Sb). The lowest unperturbed
transition energy is $\epsilon_{\pi 2d\frac 5 2}-\epsilon_{\nu
2d\frac 3 2}-\Delta M_{n-H}$, where the last term stands for
the mass difference between the neutron and the hydrogen atom,
$\Delta M_{n-H}=0.782$ MeV. If this transition energy is positive
at the mean field level - hereafter called the HF transition
energy - the system is $\beta$-stable since the CERPA correlations
could only push it up, while it is expected to be actually
$\beta$-unstable. Anticipating the discussion of the results in
Sec.~\ref{sec:results} we observe that models having positive HF
transition energies predict $\beta$-decay half-lives which are too
large in $^{132}$Sn. We therefore consider only models having a HF
energy difference $\epsilon_{\pi 2d\frac 5 2}-\epsilon_{\nu
2d\frac 3 2}<0$. This condition is indeed quite drastic, and we
found that an appreciable number of well established Skyrme models
do not fulfill it. Among these are SIII~\cite{Chabanat1997},
BSK14-17~\cite{bsk}, SKM$^*$~\cite{Chabanat1997},
SLy4-5~\cite{Chabanat1997}, SKO~\cite{sko}. In addition, the
models which predict that the HF transition energy is larger than
0.782 MeV are: RATP~\cite{Chabanat1997}, SGII~\cite{sg81},
LNS~\cite{lns}, LNS1, LNS5~\cite{lns5}, SKI1-5~\cite{ski},
SAMi~\cite{sami}. These models have therefore not been used here.

For the few remaining models, we restrict ourselves to the
parametrizations SLyIII0.7, SLyIII0.8 and
SLyIII0.9~\cite{Washiyama:2012} which predict in the bulk nuclear
matter the effective mass values $m^*/m$=0.7, 0.8 and 0.9,
respectively. We have also considered the $f_-$, $f_0$ and
$f_+$~\cite{Lesinski:2006} models which predict an effective  mass
of 0.7 in symmetric matter, with either a positive, zero, or
negative isospin splitting of the effective mass (ISEM) in neutron
matter, defined as $m^*_n/m-m^*_p/m$.
Notice that $m^*_p/m$ could be calculated in neutron
matter without any ambiguity: the proton density shall simply
be set to zero, see for instance Eq.~(\ref{eq:effmassq}).

The bulk properties of the selected interactions are given in
Table~\ref{tab:bulk}. It is observed that the saturation density
$\rho_0$, the energy per particle at saturation $E_0$, and the
symmetry energy $J_{sym}$ are very similar for these interactions.
The slope of the symmetry energy $L_{sym}$ is varying between
24.75 and 43.79~MeV, which is a rather wide range, but these models
are still considered as iso-soft ones. The incompressibility of
SLyIII0.7, SLyIII0.8 and SLyIII0.9 is quite large. However, this
does not affect the processes explored in this work. The main
difference among these models comes from their effective masses
and ISEM. The models SLyIII0.7, SLyIII0.8, SLyIII0.9
have a different effective mass in symmetric matter, and a
positive ISEM. The models $f_-$, $f_0$ and $f_+$ have the
same effective mass in symmetric matter and different signs for
the ISEM. The models SLyIII0.7, SLyIII0.8,
SLyIII0.9 and $f_+$ have a positive ISEM, as
expected from microscopic BHF and DBHF
calculations~\cite{Vandalen:2005}, while $f_0$ has no splitting
and $f_-$ has a negative ISEM. Finally, the values of the
Landau parameter for the models SLyIII0.7, SLyIII0.8 and
SLyIII0.9 is given by
\begin{equation}
G_0^{\prime}=-N_0\left[ \frac{1}{4}t_0+\frac{1}{24}t_3\rho^{\alpha_3}+\frac{1}{8}k^2_F(t_1-t_2)\right]
\end{equation}
and for the models $f_-$, $f_0$ and $f_+$,
\begin{equation}
G_0^{\prime}=-N_0\left[ \frac{1}{4}t_0+\frac{1}{4}t_3\rho^{\alpha_3}+\frac{1}{4}t_4\rho^{\alpha_4}+\frac{1}{8}k^2_F(t_1-t_2)\right]
\label{eq:gop}
\end{equation}
where $N_0 = 2k_Fm^{*}/\pi^2\hbar^2$ is the level density, with $k_F$
being the Fermi momentum and $m^{*}$ the nucleon effective mass.
In Eq.(\ref{eq:gop}), the parameter $t_4$ is coming from an additional density-dependent term
besides the usual density-dependent $t_3$
term, and the coefficient in front of the density
dependent terms have been modified w.r.t. the standard notations~\cite{Lesinski:2006}.
The values of the Landau parameter $G_0^\prime$ are given in the last column of
Table~\ref{tab:bulk}. At saturation density ($\rho=\rho_0$),
the models SLyIII0.7, SLyIII0.8, SLyIII0.9 predict rather large
values for $G_0^\prime \approx 0.3-0.35$, while the models $f_-$,
$f_0$ and $f_+$ predict smaller value with $G_0^\prime \approx 0$.
The forces $f_-$, $f_0$ and $f_+$
clearly predict not enough positive $G_0^\prime$
values~\cite{INB1}. In addition to the different effective masses
we therefore expect to observe substantial differences between
these two sets of models in the charge-exchange channel.

\begin{figure*}[t!]
\includegraphics[width=1.8\columnwidth]{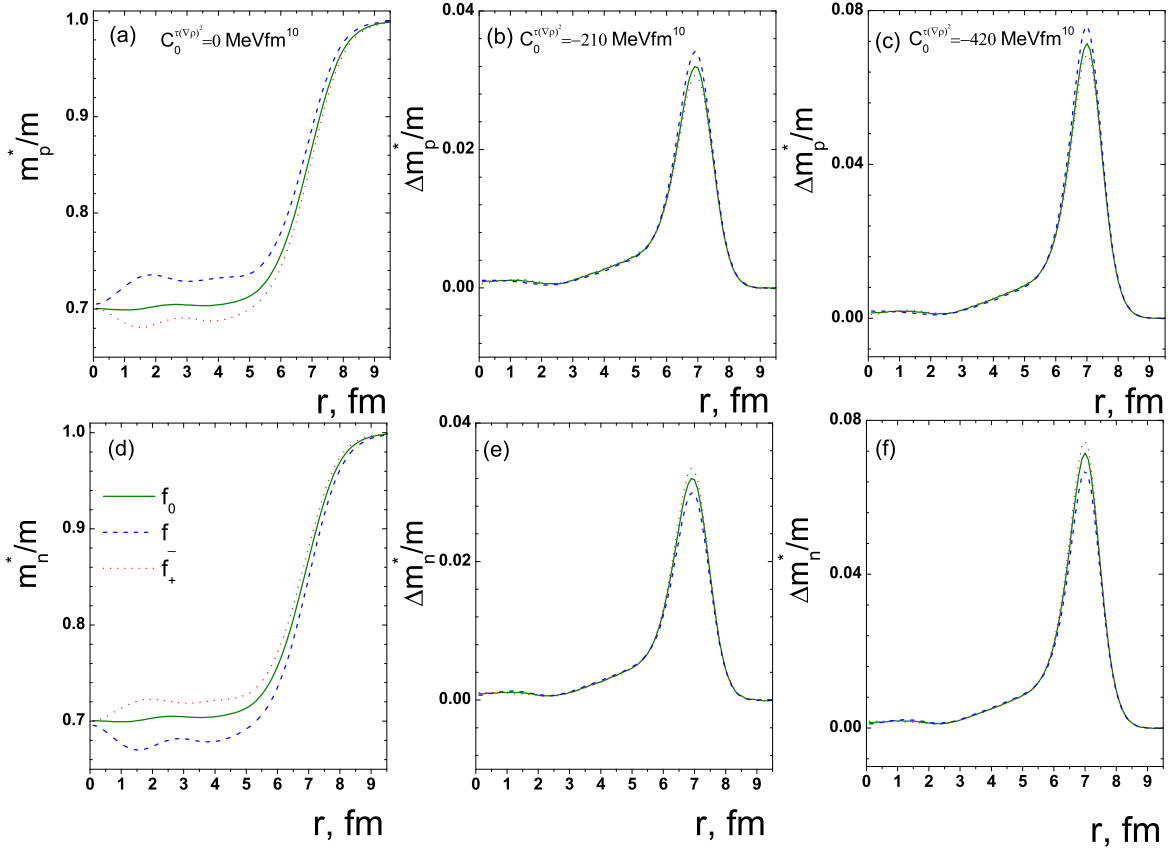}
\caption{{(Color online) Panels (a) and (d):}
Proton and neutron effective masses in $^{208}$Pb as
functions of radial distance; {panels (b), (c), (e) and (f)}:
effective-mass difference between the results of the HF
calculation with the surface peaked terms ($C^{\tau (\nabla
\rho)^2}_0$=-210 MeV fm$^{10}$ and -420 MeV fm$^{10}$), and
without. Solid, dashed and dotted lines correspond to HF
calculations with $f_0$, $f_-$ and $f_+$ forces, respectively.}
\label{fig:fig1}
\end{figure*}

\subsection{Surface-peaked effective mass correction}
In Ref.~\cite{Fantina:2011}, a surface-peaked effective mass
correction to the Skyrme-type Hamiltonian was proposed with the
following form,
\begin{equation}
\triangle\mathcal{H}=
C^{\tau (\nabla \rho)^2}_0 \tau \left(\nabla \rho \right)^2
+C^{\rho^2 (\nabla \rho)^2}_0 \rho^2 \left(\nabla \rho \right)^2,
\label{f}
\end{equation}
and the new functional can be written as $\mathcal{H}=\mathcal{H}_{sky}+\Delta\mathcal{H}$.

The first term of Eq.(\ref{f}) is designed to modify the effective
mass profile at the nuclear surface, while the second term is
introduced in order to compensate the effects of the first term in
the nuclear mean-field. Without the second term, the effects of
the first term on the mean field are too large and drastically
limit the possible values for the strength of the SPEM, as in
Ref.~\cite{Zalewski:2010}. The compensation was found to be
optimal for intermediate mass and heavy nuclei if one uses the
following constant relation between the two new
parameters~\cite{Fantina:2011}:
\begin{equation}
C^{\rho^2 (\nabla \rho)^2}_0=
\hbox{-10 fm }C^{\tau (\nabla \rho)^2}_0.
\label{ueff}
\end{equation}

{One can expect an impact of the SPEM on the properties of the lowest
quadrupole excitation if the isoscalar terms~(\ref{f}) are taken into account.
On the other hand, the energy-weighted sum rule (EWSR) is an integral characteristic
and it is particularly sensitive to the giant-resonance properties which can be
described by the EDF without the terms~(\ref{f}).
In the present work, the values of $C^{\tau (\nabla
\rho)^2}_0$=-210 and -420 MeV fm$^{10}$ are fixed so that the isoscalar
quadrupole EWSR in $^{208}$Pb is modified by
 1\% and 2\%, respectively.} Consequently, a change of less
than 0.04 of the neutron and proton effective masses at the nuclear surface of
$^{208}$Pb is predicted, see Fig.~\ref{fig:fig1}. This procedure
is slightly different from that used in Ref.~\cite{Fantina:2011},
and it leads to a SPEM less strongly peaked at the surface.
{We have added the terms~(\ref{f}) without refitting the existing
standard parameterizations. Using this perturbative approach, we observe
a small change of the binding energies which is larger than the tolerance of the protocol
for the parameter fitting. In particular, in $^{208}$Pb the binding energy
changes by 0.35\% for the SLyIII0.9 set, 0.37\% for the SLyIII0.8 set,
0.38\% for the SLyIII0.7 set and 0.45\% for the $f_0$, $f_-$ and $f_+$ sets.
A fine tuning of other parameters in order to compensate for these energy
changes has still to be done.}

In Fig.~\ref{fig:fig1} are shown the effective mass profiles in
$^{208}$Pb for the $f_0$, $f_-$ and $f_+$ models where we have
considered different values of the parameter governing the
strength of the SPEM, $C^{\tau (\nabla
\rho)^2}_0$=0, -210 and -420 MeV fm$^{10}$. We remind that the
differences between the models $f_0$, $f_-$ and $f_+$ are mostly
the ISEM in asymmetric matter:
$f_+$ has $m^*_n>m^*_p$ in neutron rich matter, while $f_-$ has
$m^*_n<m^*_p$, and $f_0$ has $m^*_n=m^*_p$ in the same conditions
of isospin asymmetry. The effect of the sign difference of the
effective mass splitting can also be observed on {panels (a) and (d)}
(without SPEM): Since $^{208}$Pb is a neutron rich nucleus, the
neutron effective mass is larger than the proton one for the f$_+$
model, an opposite effect is found for f$_-$, and no effect is
observed for $f_0$. Additionally, it is observed in {panels (b), (c), (e) and (f)}
that the SPEM correction is almost unaffected by the
effective mass splitting, since the correction is isoscalar.

\subsection{Calculations of $\beta$-decay rates}
We describe the collective modes in the charge-exchange random
phase approximation (CERPA) using the same Skyrme interactions as
above. Making use of the finite-rank separable approximation
(FRSA)~\cite{gsv98,svg08,svg12} for the p-h interaction enables us
to perform CERPA calculations in very large configuration spaces.
Although it is well known that the tensor interaction influences
also the description of the $\beta^{-}$-decay
half-lives~\cite{mb13}, in the present study the tensor force is
neglected in order to focus on the impact of the SPEM.

The experimentally known values of the half-lives put an indirect
constraint on the calculated GT strength distributions within the
$Q_{\beta}$-window. To calculate the half-lives an approximation
worked out in Ref.~\cite{EBDNS99} is used. It allows one to avoid
an implicit calculation of the nuclear masses and
$Q_{\beta}$-values. However, one should realize that the related
uncertainty in constraining the parent nucleus ground state
calculated with the chosen Skyrme interaction is transferred to
the values of the neutron and proton chemical potentials. In the
allowed GT approximation, the $\beta^{\pm}$-decay rate is
expressed by summing the probabilities of the energetically
allowed transitions (in units of $G_{A}^{2}/4\pi$) weighted with
the integrated Fermi function. For the $\beta^{-}$-decay case we
have:
\begin{equation}
T^{\beta^{-}}_{1/2} =\frac{D}{\left(\frac{G_{A}}{G_{V}}\right)^{2}
\sum_{k}f_{0}(Z+1,A,E_i-E_{1^+_k})B(GT)^{-}_{k}},
\label{eq:halflife}
\end{equation}
\begin{equation}
E_i-E_{1^+_k}\approx\Delta M_{n-H}+\mu_n-\mu_p-E_k
\end{equation}
while for the $\beta^{+}$-decay case this becomes:
\begin{equation}
T^{\beta^{+}}_{1/2} =\frac{D}{\left(\frac{G_{A}}{G_{V}}\right)^{2}
\sum_{k}f_{0}(-Z+1,A,E_i-E_{1^+_k})B(GT)^{+}_{k}},
\label{eq:halflife+}
\end{equation}
\begin{equation}
E_i-E_{1^+_k}\approx-\Delta M_{n-H}-2m_{e}-\mu_n+\mu_p-E_k.
\end{equation}
Here, $D$=6147~\cite{Suhonen} is a constant,
$G_A/G_V$=1.25~\cite{Suhonen} is the ratio of the weak
axial-vector and vector coupling constants, $m_{e}$ is the
positron mass; $\mu_n$ and $\mu_p$ are the neutron and proton
chemical potentials. $E_i$ is the ground state energy of the
parent nucleus $(Z,A)$ and $E_{1_k^+}$ denotes a state of the daughter
nucleus. The $E_k$ are the {$1^+$} eigenvalues of the CERPA
equations. The CERPA wave functions allow us to determine GT
transitions whose operator is $\hat{O}_{\pm} = \sum_{i, m} t_{\pm}(i) \sigma_m(i)$.
\begin{equation}
B(GT)^{\pm}_{k} = \left|\langle N\pm 1, Z\mp 1; 1_k^+
|\hat{O}^{\pm}| N,Z; 0_{gs}^+ \rangle\right|^2 .
\label{eq:bgt}
\end{equation}

Expressions (\ref{eq:halflife})-(\ref{eq:bgt}) will be used in the
next section to calculate the $\beta$-decay rates and the
collective modes. All the calculations are performed without any
quenching factor.
\begin{table}[tb]
\caption{Energy differences between the dominant s.p. states in
$^{132}$Sn and $^{100}$Sn. For each Skyrme parameterization, the
energy difference is calculated with the surface peaked term or
without ($C^{\tau (\nabla \rho)^2}_0=0$). See text for
more details.} \label{tab:spe}
\begin{ruledtabular}
\begin{tabular}{ccccc}
 & & $^{132}$Sn & $^{100}$Sn & $^{78}$Ni\\
 Skyrme  &$C^{\tau (\nabla \rho)^2}_0$ &$\epsilon_{\pi 2d\frac 5 2}-$&$\epsilon_{\nu 1g\frac 7 2}-$&$\epsilon_{\pi 2p\frac 3 2}-$\\
               &&$\epsilon_{\nu 2d\frac 3 2}$&$\epsilon_{\pi 1g\frac 9 2}$&$\epsilon_{\nu 2p\frac 1 2}$\\
   & (MeV fm$^{10}$) & (MeV) & (MeV) & (MeV)\\
\noalign{\smallskip}\hline\noalign{\smallskip}
  SLyIII0.7    &        0                &      0.3   &    -7.2   &-5.1      \\
  SLyIII0.7    &     -210                &      0.2   &    -7.3   &-5.2      \\
  SLyIII0.7    &     -420                &      0.1   &    -7.3   &-5.3      \\
  SLyIII0.8    &        0                &     -0.6   &    -7.5   &-6.2      \\
  SLyIII0.8    &     -210                &     -0.7   &    -7.5   &-6.3      \\
  SLyIII0.8    &     -420                &     -0.9   &    -7.6   &-6.4      \\
  SLyIII0.9    &        0                &     -1.3   &    -7.7   &-7.0      \\
  SLyIII0.9    &     -210                &     -1.4   &    -7.7   &-7.2      \\
  SLyIII0.9    &     -420                &     -1.6   &    -7.8   &-7.3      \\
   $f_+$       &        0                &     -0.6   &    -5.9   &-5.8      \\
   $f_+$       &     -210                &     -0.7   &    -6.0   &-5.9      \\
   $f_+$       &     -420                &     -0.8   &    -6.2   &-6.0      \\
   $f_0$       &        0                &     -0.5   &    -6.0   &-5.6      \\
   $f_0$       &     -210                &     -0.6   &    -6.1   &-5.7      \\
   $f_0$       &     -420                &     -0.7   &    -6.2   &-5.8      \\
   $f_-$       &        0                &     -0.4   &    -6.2   &-5.4      \\
   $f_-$       &     -210                &     -0.5   &    -6.3   &-5.5      \\
   $f_-$       &     -420                &     -0.6   &    -6.5   &-5.6      \\
\end{tabular}
\end{ruledtabular}
\end{table}
%

%
%==============================================================
%
\section{Results for collective modes and $\beta$-decay rates}
\label{sec:results}
We now analyze first the results of the $\beta$-decay rates which
are sensitive to the low-energy part of the CERPA strength, and
then the GT collective modes. The effects of the SPEM will be
discussed.

\begin{table}
\caption{{SPEM effects on $\beta^-$-decay properties of $^{132}$Sn.
Data are from Ref.~\cite{nndc}.}}
\label{tab:res1}
\begin{ruledtabular}
\begin{tabular}{ccccc}
 Skyrme&$C^{\tau (\nabla \rho)^2}_0$&{$E_i-E_{1^+_1}$}&{$B(GT)^{-}_{1}$}&$T_{1/2}$\\
       &(MeV fm$^{10}$)             &(MeV)          &               &(s)      \\
\noalign{\smallskip}\hline\noalign{\smallskip}
SLyIII0.7    &                          0  &0.07&2.6                & 389400  \\
SLyIII0.7    &                       -210  &0.21&2.7                &   9840  \\
SLyIII0.7    &                       -420  &0.34&2.7                &   1930  \\
SLyIII0.8    &                          0  &0.97&2.5                &     57  \\
SLyIII0.8    &                       -210  &1.11&2.5                &     33  \\
SLyIII0.8    &                       -420  &1.26&2.6                &     21  \\
SLyIII0.9    &                          0  &1.70&2.4                &    6.7  \\
SLyIII0.9    &                       -210  &1.84&2.5                &    4.7  \\
SLyIII0.9    &                       -420  &2.01&2.6                &    3.3  \\
$f_+$        &                          0  &1.12&4.6                &     18  \\
$f_+$        &                       -210  &1.25&4.6                &     12  \\
$f_+$        &                       -420  &1.36&4.6                &    8.5  \\
$f_0$        &                          0  &1.14&5.9                &     13  \\
$f_0$        &                       -210  &1.27&5.8                &    8.8  \\
$f_0$        &                       -420  &1.37&5.8                &    6.5  \\
$f_-$        &                          0  &1.12&8.8                &    6.4  \\
$f_-$        &                       -210  &1.25&8.7                &    5.0  \\
$f_-$        &                       -420  &1.36&8.6                &    3.6  \\
Expt.        &                             &1.794$\pm$0.009&           &39.7$\pm$0.8  \\
\end{tabular}
\end{ruledtabular}
\end{table}
\begin{table}
\caption{{SPEM effects on $\beta^-$-decay properties of $^{78}$Ni.
Data are from Refs.~\cite{78Ni}.}}
\label{tab:res1p}
\begin{ruledtabular}
\begin{tabular}{ccccc}
 Skyrme&$C^{\tau (\nabla \rho)^2}_0$&{$E_i-E_{1^+_1}$}&{$B(GT)^{-}_{1}$}&$T_{1/2}$\\
       &(MeV fm$^{10}$)             &(MeV)          &               &(s)      \\
\noalign{\smallskip}\hline\noalign{\smallskip}
  SLyIII0.7    &                          0  &5.49&1.0&     0.157    \\
  SLyIII0.7    &                       -210  &5.61&1.1&     0.140    \\
  SLyIII0.7    &                       -420  &5.74&1.1&     0.121    \\
  SLyIII0.8    &                          0  &6.60&1.0&     0.057    \\
  SLyIII0.8    &                       -210  &6.73&1.0&     0.051    \\
  SLyIII0.8    &                       -420  &6.87&1.0&     0.045    \\
  SLyIII0.9    &                          0  &7.48&1.0&     0.025    \\
  SLyIII0.9    &                       -210  &7.61&1.0&     0.023    \\
  SLyIII0.9    &                       -420  &7.79&1.0&     0.020    \\
    $f_+$      &                          0  &6.40&1.9&     0.031    \\
    $f_+$      &                       -210  &6.51&1.9&     0.028    \\
    $f_+$      &                       -420  &6.60&1.9&     0.027    \\
    $f_0$      &                          0  &6.33&2.6&     0.020    \\
    $f_0$      &                       -210  &6.44&2.5&     0.019    \\
    $f_0$      &                       -420  &6.52&2.5&     0.018    \\
    $f_-$      &                          0  &6.33&3.9&     0.010    \\
    $f_-$      &                       -210  &6.42&3.9&     0.009    \\
    $f_-$      &                       -420  &6.50&3.8&     0.009    \\
Expt.          &                             &    &   &0.1222$\pm$0.0051\\
\end{tabular}
\end{ruledtabular}
\end{table}

The p-h interaction in the spin-isospin channel is assumed of the
following form:
\begin{equation}
V({\bf r}_1,{\bf r}_2) = N_0^{-1} G_0^{'}(r_1){\bo \sigma
}^{(1)}\cdot{\bo \sigma}^{(2)} {\bo \tau }^{(1)}\cdot{\bo \tau
}^{(2)} \delta ({\bf r}_1 - {\bf r}_2)
\end{equation}
where ${\bo \sigma}^{(i)}$ and ${\bo \tau}^{(i)}$ are the spin and
isospin operators. As expected, the largest contribution
to the calculated $\beta^{\pm}$-decay half-life comes
from the $1_1^+$ state, the structure of which is dominated by  one
unperturbed configuration. They are the 1p-1h configurations
$\{\pi 2d\frac 5 2,\nu 2d\frac 3 2\}$, $\{\nu 1g\frac 7 2, \pi
1g\frac 9 2\}$ and {$\{\pi 2p\frac 3 2,\nu 2p\frac 1 2\}$} of
$^{132}$Sn, $^{100}$Sn and $^{78}$Ni, respectively. In other
words, the $1_1^+$ state is non-collective and, therefore, the
$\beta$-decay is related to the lowest unperturbed $1^+$ energy.
We first examine the s.p. energy differences given in
Table~\ref{tab:spe} for the selected Skyrme models and for various
strength of the SPEM parameter $C^{\tau (\nabla \rho)^2}_0$. They
are small (about 1 MeV) in $^{132}$Sn but rather large in
$^{78}$Ni and $^{100}$Sn (5 to 8 MeV).
In $^{100}$Sn, the energy differences without SPEM are mostly sensitive to the Coulomb
component of the EDF, with a small additional effect due to the effective mass
(the larger effective mass, the smaller the energy difference).
{In $^{132}$Sn, the energy difference is related mostly to the
symmetry energy: the larger the symmetry energy, going from SLyIII0.9 to SLyIII0.7 for instance,
the larger the energy difference.}
In addition, the increase of the effective mass also contributes, with a
smaller impact, to the decrease the energy difference,
{as it can be deduced from the comparison of the energy difference for the forces
$f_-$, $f_0$, and $f_+$ which has increasing effective mass in neutron rich matter,
see Fig.~\ref{fig:fig1}.}
It can be seen (cf. Table~\ref{tab:spe}) that the shifts in the energy differences
between the cases without, and with maximal SPEM ($C^{\tau
(\nabla \rho)^2}_0$=-420~MeV~fm$^{10}$) are almost constant and
independent of the models considered. It varies by about 0.3 MeV
in $^{132}$Sn and in $^{78}$Ni. From Table II, we can anticipate
that the SPEM will have a larger impact on the calculation of the
$\beta$ half-life of $^{132}$Sn and a weaker one in the case of
$^{78}$Ni and $^{100}$Sn. In $^{132}$Sn, the experimental value is
-1.305~MeV~\cite{sps132Sn}. No data exist for $^{78}$Ni and
$^{100}$Sn.

\begin{table}
\caption{{SPEM effects on $\beta^+$-decay properties of $^{100}$Sn.
Data are from Ref.~\cite{100Sn}.}}  \label{tab:res2}
\begin{ruledtabular}
\begin{tabular}{ccccc}
  Skyrme       &$C^{\tau (\nabla \rho)^2}_0$&{$E_i-E_{1^+_1}$}&{$B(GT)^{+}_{1}$}&$T_{1/2}$\\
               &(MeV fm$^{10}$)             &(MeV)          &               &(s)      \\
\noalign{\smallskip}\hline\noalign{\smallskip}
  SLyIII0.7    &                          0 & 4.26 & 15.2 & 0.232 \\
  SLyIII0.7    &                       -210 & 4.38 & 15.2 & 0.221 \\
  SLyIII0.7    &                       -420 & 4.42 & 15.2 & 0.213 \\
  SLyIII0.8    &                          0 & 4.60 & 15.1 & 0.178 \\
  SLyIII0.8    &                       -210 & 4.64 & 15.1 & 0.172 \\
  SLyIII0.8    &                       -420 & 4.67 & 15.0 & 0.167 \\
  SLyIII0.9    &                          0 & 4.86 & 15.1 & 0.138 \\
  SLyIII0.9    &                       -210 & 4.90 & 15.0 & 0.134 \\
  SLyIII0.9    &                       -420 & 4.92 & 15.0 & 0.131 \\
    $f_+$      &                          0 & 3.47 & 16.3 & 0.593 \\
    $f_+$      &                       -210 & 3.62 & 16.3 & 0.492 \\
    $f_+$      &                       -420 & 3.72 & 16.2 & 0.433 \\
    $f_0$      &                          0 & 3.80 & 16.8 & 0.381 \\
    $f_0$      &                       -210 & 3.94 & 16.8 & 0.323 \\
    $f_0$      &                       -420 & 4.04 & 16.7 & 0.290 \\
    $f_-$      &                          0 & 4.38 & 17.6 & 0.190 \\
    $f_-$      &                       -210 & 4.51 & 17.5 & 0.168 \\
    $f_-$      &                       -420 & 4.60 & 17.5 & 0.154 \\
    Expt.      &                            & 3.08$\pm$0.34& &    1.16$\pm$0.20\\
\end{tabular}
\end{ruledtabular}
\end{table}

{The $E_i-E_{1^+_1}$ energies, the $B(GT)^{-}_{1}$ values and $\beta^-$-decay half-lives of $^{132}$Sn and $^{78}$Ni are
given in Tables~\ref{tab:res1} and \ref{tab:res1p}, the $\beta^+$-decay properties of $^{100}$Sn in Table~\ref{tab:res2}.
The evolution of the transition energies and the $B(GT)^{\pm}_{1}$ values is reflected in the half-life behaviour,
see Eqs.(\ref{eq:halflife}) and (\ref{eq:halflife+}).}
As in Table~\ref{tab:spe},
the results shown in {Tables~\ref{tab:res1}, \ref{tab:res1p} and \ref{tab:res2}}
correspond to the selected interactions with and without the SPEM
represented by the value of the parameter $C^{\tau (\nabla
\rho)^2}_0$. In the case of $^{132}$Sn, the model SLyIII0.7
predicts positive energy differences for the dominant transition
of the $\beta$-decay half-lives (see Table~\ref{tab:spe}), and it
leads to half-lives which are much larger than the experimental
value, as anticipated.

One can see from {Tables~\ref{tab:res1}, \ref{tab:res1p} and \ref{tab:res2}} that the
$\beta$-decay half-lives are much more sensitive to the effective
mass distribution in the case of {the low-$Q_{\beta}$ nucleus} 
$^{132}$Sn than in $^{100}$Sn and
$^{78}$Ni. For $^{132}$Sn, a strong decrease of the half-life can
be directly correlated to either the increase of the effective
mass in symmetric matter, or to the increase of the SPEM, while in
$^{78}$Ni and $^{100}$Sn, the correlation, while still being
present, is much less pronounced. This can be easily understood
from the energy difference of the most important transition given
in Table~\ref{tab:spe}: The energy differences are much smaller in
the case of $^{132}$Sn than in the case of $^{100}$Sn and
$^{78}$Ni, which makes the $\beta$-decay half-lives more sensitive
to a small modification of the s.p. energies induced by the SPEM.

\begin{figure}[t!]
\includegraphics[width=1.0\columnwidth]{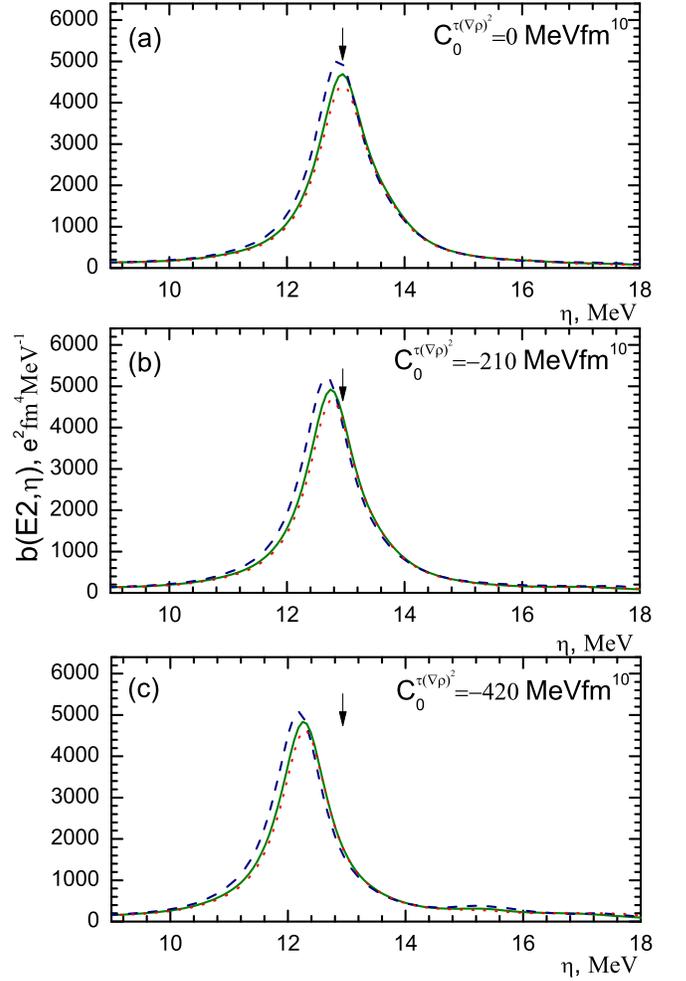}
\caption{{(Color online)} The quadrupole strength distribution of
$^{208}$Pb. Solid, dashed and dotted lines correspond to RPA
calculations with $f_0$, $f_-$ and $f_+$ models, respectively. The
experimental centroid of the GQR is at $10.89\pm0.30$
MeV~\cite{GQR208Pb}.} \label{fig:fig2}
\end{figure}
\begin{table}
\caption{{SPEM effects on the energy and B(E2)-value for the up-transition
to the first $2^{+}$ state in $^{208}$Pb.
Data are from Ref.~\cite{nndc}.}}  \label{tab:res3}
\begin{ruledtabular}
\begin{tabular}{cccc}
Skyrme  &$C^{\tau (\nabla \rho)^2}_0$ &Energy &$B(E2;0_{gs}^+\rightarrow 2_1^+)$ \\
        &(MeV fm$^{10}$)              &(MeV)  &(e$^2$fm$^4$)                     \\
\noalign{\smallskip}\hline\noalign{\smallskip}
$f_+$   &          0                  &5.12   &3130                              \\
$f_+$   &       -210                  &5.09   &2530                              \\
$f_+$   &       -420                  &5.09   &2180                              \\
$f_0$   &          0                  &5.13   &3250                              \\
$f_0$   &       -210                  &5.09   &2650                              \\
$f_0$   &       -420                  &5.09   &2310                              \\
$f_-$   &          0                  &5.09   &3440                              \\
$f_-$   &       -210                  &5.06   &2850                              \\
$f_-$   &       -420                  &5.06   &2500                              \\
Expt.   &                             &4.09   &3000$\pm$300                      \\
\end{tabular}
\end{ruledtabular}
\end{table}
\begin{figure}[t!]
\includegraphics[width=1.0\columnwidth]{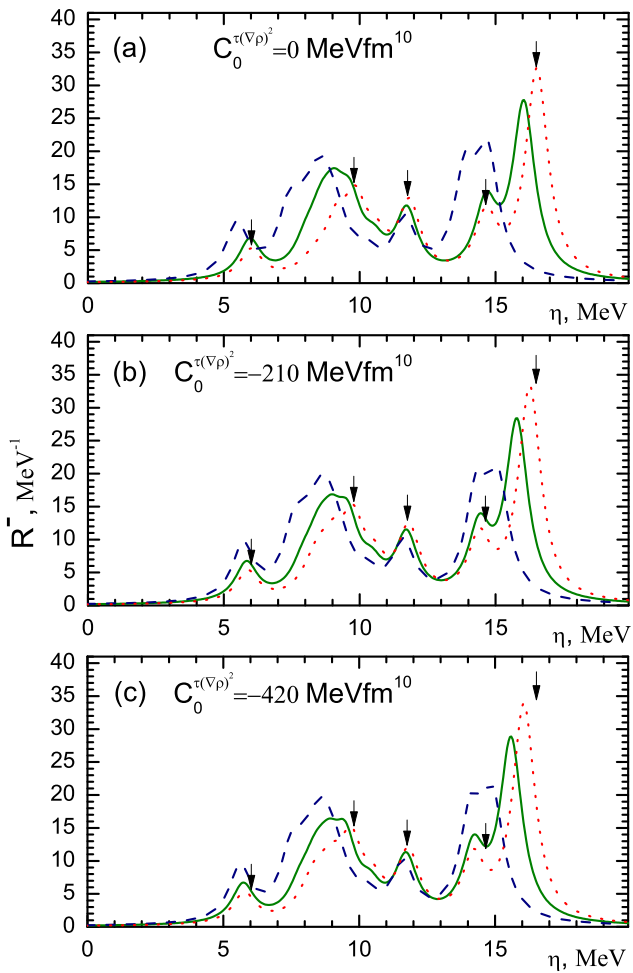}
\caption{{(Color online)} Same as Fig.~\ref{fig:fig2}, for the GT strength
distribution obtained within the CERPA. The experimental centroids
of the GTR is at 19.2 MeV~\cite{GTR208Pb1,GTR208Pb2}.}
\label{fig:fig3}
\end{figure}

Let us examine whether the SPEM could improve the agreement
between the model predictions and the experimental values.
{As one can see from Tables III, IV and V the inclusion of
the terms~(\ref{f}) leads to minor effects on the $B(GT)^{\pm}_{1}$ values.
We find that the SPEM induces an increase of the transition energies
and it results in a decrease of the half-lives.}
We first concentrate on the models SLyIII0.7, SLyIII0.8 and
SLyIII0.9, which correspond to different values of the effective
mass in symmetric matter. From the comparison of the theoretical
predictions with the experimental {half-lives} shown in
{Tables~\ref{tab:res1}, \ref{tab:res1p} and \ref{tab:res2}},
it is difficult to conclude which model is better: For $^{132}$Sn,
the model SLyIII0.8 is preferred,
for $^{78}$Ni and $^{100}$Sn, it is SLyIII0.7.
Now, if we concentrate on the models
f$_+$, f$_0$ and f$_-$, it is f$_+$ which always comes the closest
to the experimental value. This indicates that, in addition to the
effective mass, the residual interaction is very important. It was
already anticipated that the value of the Landau parameter
$G_0^\prime$ for the selected models (cf. Table~\ref{tab:bulk})
could have an impact on charge-exchange related observables. For
the models f$_+$, f$_0$ and f$_-$ the values of $G_0^\prime$ are
too small. Since the impact of the SPEM on the $\beta$-decay rates
in $^{78}$Ni and $^{100}$Sn is quite small, these nuclei could be,
in the future, used to calibrate the residual interaction almost
independently from the profile of the effective mass. The
modification of $G_0^\prime$ could be obtained from a refitting of
the Skyrme functional with a different value of the strength of
the SPEM. One could increase $G_0^\prime$ by about 0.1-0.2 by
introducing the spin-density dependent extension of the Skyrme
model~\cite{Margueron:2009a,Margueron:2009b}. This will be left
for future investigations.

Up to this point, we have mostly focused on the relation between
the SPEM and the low energy part of the strength, since it
represents the main contribution to the $\beta$-decay rates. We
now turn to the higher energy part and show in
Figs.~\ref{fig:fig2} and \ref{fig:fig3} the effects of the SPEM on
the properties of the GQR and GTR in $^{208}$Pb. In the figures,
the calculated strength distributions are folded out with a
Lorentzian distribution of 1 MeV width. The excitation energies
refer to the ground state of the parent nucleus $^{208}$Pb. The
arrows indicate the maxima of the strength distributions
corresponding to the case of the $f_+$ model and $C^{\tau (\nabla
\rho)^2}_0$=0 MeV fm$^{10}$. Since the isoscalar quadrupole
{EWSR} is changed by only about 1\% by the SPEM,
we expect the collective modes at higher energy to be only
marginally impacted.

{The GQR strength distribution consists mostly of a main peak. Comparing
the cases without SPEM and with maximal SPEM, we find that the
peak is shifted down by about 500~keV. One can notice that the GQR
strength distribution is almost identical for the three models
$f_0$, $f_+$ and $f_-$. As can be seen from Table~\ref{tab:res3},
while the $2_1^{+}$ energy is practically unaffected by the SPEM,
the B(E2) value is decreasing as the SPEM increases.
%the properties of the $2_1^{+}$ state is more sensitive to the
%inclusion of the terms~(\ref{f}). The SPEM effects produce a
%sizable impact on the B(E2) value which is reduced by a factor
%1.4. The $2_1^{+}$ energy is practically unchanged.
Some overestimate of the experimental energy indicates that there is
room for the two-phonon effects~\cite{svg04}.}

The GTR is much more fragmented than the GQR, as seen in
Fig.~\ref{fig:fig3}. The strength distribution is globally shifted up
as the isospin splitting is going from positive (f$_+$) to
negative values (f$_-$). As in the case of the GQR, the high
energy peaks of the strength distribution are shifted to lower
energies (by about 500 keV) as the SPEM gets larger. This is an
effect of the slight increase of the level density induced by the
SPEM.

%
%=====================================================================
%
\section{Conclusions}

Starting from different Skyrme EDF which predict
$^{132}$Sn $\beta$-unstable, we have studied the effects of
introducing a surface-peaked effective mass on top of existing
Skyrme models. The main effect of this additional term is a
compression of the s.p. level spacing around the Fermi level, or
equivalently, an increase of the level density. This
systematically increases the $\beta$-decay rates (i.e., decreases
the half-lives). The collective modes at higher energy are only
slightly impacted by the SPEM.

This work is a first step towards improving Skyrme functionals by
adding extra terms to the energy functional. Our motivation is
based on both having a better agreement with nuclear data, and
also predicting weak transition rates for astrophysical
applications. The results of our analysis allow for a better
understanding of the effects at play. The $\beta$-decay rates in
doubly-magic unstable nuclei ($^{100,132}$Sn, $^{78}$Ni) are
indeed very sensitive both to the s.p. energies and residual
interactions, and none of the Skyrme models selected in this work
are fully satisfactory in this respect. From our analysis, we have
however identified two nuclei ($^{100}$Sn, $^{78}$Ni) where the
$\beta$-decay half-lives are only weakly impacted by the SPEM.
They can be considered as good benchmark nuclei since they
potentially offer the possibility to calibrate the residual
interaction, with a weak influence of the effective mass. In a
complementary approach, $^{132}$Sn could be used to test different
strengths of the SPEM, for a fixed residual interaction.

The tensor force has not been considered in this work, although it
can affect the neutron-proton s.p. energies in some cases. We have
aimed at understanding just the contribution of the SPEM to the
$\beta$-decay and GT mode in order to disentangle the respective
roles of the effective mass and the residual interaction. An
additional modification of the Skyrme functional was proposed
earlier in order to stabilize the nuclear matter equation of
state~\cite{Margueron:2009a,Margueron:2009b}. It has been recently
used in nuclei and, since it brings an additional repulsive term
to the $G_0^\prime$ Landau parameter, it was shown to shift the
centroids of the GT collective mode to higher energies by a few
hundreds keV up to 1 MeV~\cite{Wen:2014}. In the future, we plan to
explore the predictions of a general mean field model including all
these ingredients, and to compare them to known experimental data,
as done in this work. These calibration processes are important to
set-up boundaries for the additional parameters before making
predictions for astrophysical cases.

%
%==================================================================
%
\section*{Acknowledgments}
A.P.S. {and I.N.B.} thank the hospitality of IPN-Orsay
and IPN-Lyon where a part of this work was done. This work is
partly supported by the IN2P3-JINR agreement and the ANR project
"SN2NS: supernova explosions, from stellar core- collapse to neutron
stars and black holes".
%
%==================================================================
%

\appendix

\section{Decomposition of the Skyrme functional into neutron and proton channels}
\label{app:npc} Here, the Skyrme functional is expressed in terms
of the neutron and proton densities instead of the isoscalar and
isovector densities,
\begin{eqnarray}
\mathcal{H}_{sky}(r) =  \sum_{q=n,p} h_q^\rho + h_q^\nabla + h_q^{J} ,
\label{eq:functional}
\end{eqnarray}
where the different terms of the energy density are
\begin{eqnarray}
h_q^\rho &=& \frac{\hbar^2}{2m} f_q^\mathrm{Sky} \tau_q+ (C_0^\rho+C_1^\rho) \rho_q^2 +
(C_0^\rho-C_1^\rho) \rho_q\rho_{\bar q} \label{eq:ssky} \nonumber \\  \\
h_q^\nabla &=& -(C_0^{\Delta \rho}+C_1^{\Delta \rho}) (\nabla \rho_q)^2
- (C_0^{\Delta \rho}-C_1^{\Delta \rho}) \nabla \rho_q \cdot \nabla \rho_{\bar q}\nonumber \\ \\
h_q^{J} &=&\frac{1}{2}(C_0^J+C_1^J) J_q^2+\frac{1}{2}(C_0^J-C_1^J) J_q J_{\bar q} \nonumber \\
&&-\left[(C_0^{\nabla J}+C_1^{\nabla J}) \nabla \rho_q+(C_0^{\nabla J}-C_1^{\nabla J})\nabla \rho_{\bar q}\right]\cdot J_q ,
\nonumber \\
\end{eqnarray}
and the effective mass factor $f_q^\mathrm{Sky}=m/m_q^*$ is defined as
\begin{equation}
f_q^\mathrm{Sky} = 1 + \frac{2m}{\hbar^2}[(C_0^\tau+C_1^\tau)\rho_q+ (C_0^\tau-C_1^\tau)\rho_{\bar q} ] .
\end{equation}

By functional derivation the one-body Hamiltonian $\mathcal{H}_q$
is obtained as,
\begin{eqnarray}
\mathcal{H}_q &=& -\frac{\hbar^2}{2m} \nabla \cdot f_q^\mathrm{Sky}(r) \nabla + V_q(r) -\nonumber\\
&&\frac{i}{2}\sum_{\sigma'}[W_q\cdot(\nabla\times\langle\sigma\vert \sigma\vert\sigma'\rangle)
+(\nabla\times\langle\sigma\vert \sigma\vert\sigma'\rangle)\cdot W_q]\nonumber\\
\end{eqnarray}
where the central potential is given by
\begin{equation}
V_q^\mathrm{Sky}(r) = V_q^\rho(r) + V_q^\nabla(r) + V_q^J(r).
\label{eq:vqsky}
\end{equation}
Here, the central-density potential is given by:
\begin{eqnarray}
V_q^\rho(r) &=& (C_0^\tau+C_1^\tau)\tau_q+(C_0^\tau-C_1^\tau)\tau_{\bar q} \nonumber \\
&&+2[(C_0^\rho+C_1^\rho)\rho_q+(C_0^\rho-C_1^\rho)\rho_{\bar q}]\nonumber \\
&&+\frac{\partial}{\partial\rho_0}(C_0^\rho+C_1^\rho)\rho_q^2+
\frac{\partial}{\partial\rho_0}(C_0^\rho-C_1^\rho)\rho_q\rho_{\bar q}] ,\nonumber \\
\end{eqnarray}
the central-gradient potential by:
\begin{eqnarray}
V_q^\nabla(r) = 2(C_0^{\Delta\rho}+C_1^{\Delta\rho})\nabla^2\rho_q
+ 2(C_0^{\Delta\rho}-C_1^{\Delta\rho})\nabla^2\rho_{\bar q} ,\nonumber \\
\end{eqnarray}
and the central-$J$ potential by:
\begin{eqnarray}
V_q^J(r) = (C_0^{\nabla J}+C_1^{\nabla J}) \nabla\cdot J_q+
(C_0^{\nabla J}-C_1^{\nabla J}) \nabla\cdot J_{\bar q}.\nonumber \\
\end{eqnarray}

The spin-orbit potential is:
\begin{eqnarray}
W_q(r) &=& -(C_0^{\nabla J}+C_1^{\nabla J})\nabla\rho_q
-(C_0^{\nabla J}-C_1^{\nabla J})\nabla\rho_{\bar q} \nonumber \\
&&+(C_0^{J}+C_1^{J})J_q+(C_0^{J}-C_1^{J})J_{\bar q} .
\label{eq:sopotential}
\end{eqnarray}

\section{Modification of the mean-field equations induced by the SPEM}
The kinetic energy correction induced by the effective mass in
Eq.~(\ref{eq:ssky}) is now given by
$f_q=f_q^\mathrm{Sky}+f_q^\mathrm{corr}$ where
\begin{eqnarray}
f_q^\mathrm{corr} =  \frac{2m}{\hbar^2} C^{\tau (\nabla \rho)^2}_0 \left(\nabla \rho(\mathbf{r})\right)^2 \ ,
\end{eqnarray}
and the mean field central potential~(\ref{eq:vqsky}) reads:
\begin{equation}
V_q(\mathbf{r}) = V_q^\mathrm{Sky}(\mathbf{r}) + V^\mathrm{corr}(\mathbf{r}) \ ,
\end{equation}
where $V_q^\mathrm{Sky}(\mathbf{r})$ is the mean field deduced from the Skyrme interaction, e.g. Eq.~(\ref{eq:vqsky}),
and $V^\mathrm{corr}(\mathbf{r})$ is the correction term induced by Eq.~(\ref{f}):
\begin{eqnarray}
V^\mathrm{corr}(\mathbf{r}) = -2 C^{\tau (\nabla \rho)^2}_0 \Big( \tau(\mathbf{r}) \nabla^2 \rho(\mathbf{r})
+ \nabla \tau(\mathbf{r}) \nabla \rho(\mathbf{r}) \Big) \nonumber \\
- 2 C^{\rho^2 (\nabla \rho)^2}_0 \Big( \rho(\mathbf{r}) (\nabla \rho(\mathbf{r}))^2
+ \rho(\mathbf{r})^2 \nabla^2 \rho(\mathbf{r}) \Big) \; . \nonumber \\
\label{chap:effmass:eq:ucorr}
\end{eqnarray}

%
%==================================================================
%
\end{document}